\documentstyle[12pt,epsf,worldsci]{article}
\font\tenrm=cmr10

\def\vn{{\vec{{\rm n}}}}                    
\def\vri{{\vec{{\rm r}}}_i}                 
\def\vrj{{\vec{{\rm r}}}_j}                 
\def\vrij{{\vec{{\rm r}}}_{ij}}             
\def\vrr{{\vec{{\rm r}}}_{{\rm r}}}         
\def\vrg{{\vec{{\rm r}}}_{{\rm g}}}         
\def\vrb{{\vec{{\rm r}}}_{{\rm b}}}         
\def\rrg{{{\rm r}}_{{\rm rg}}}              
\def\rgb{{{\rm r}}_{{\rm gb}}}              
\def\rbr{{{\rm r}}_{{\rm br}}}              
\def\chic{\mbox{\raisebox{1mm}{$\chi$}}_{\rm Correlation}}
\begin{document}
\renewenvironment{thebibliography}[1]
  { \begin{list}{\arabic{enumi}.}
    {\usecounter{enumi} \setlength{\parsep}{0pt}
     \setlength{\itemsep}{3pt} \settowidth{\labelwidth}{#1.}
     \sloppy
    }}{\end{list}}

\parindent=1.5pc

\begin{center}{{\bf SU(3) String-Flip Potential Models and Nuclear Matter}\\
\vglue 1.0cm
{M. BOYCE\footnote{Talk presented by M. Boyce.} and P.J.S. WATSON}\\
\baselineskip=14pt
{\it Ottawa-Carleton Institute for Physics, Carleton University,}\\
\baselineskip=14pt
{\it  Ottawa, Ontario, Canada, K1S-5B6}\\
\vglue 0.8cm
{\tenrm ABSTRACT}}
\end{center}
{\rightskip=3pc
 \leftskip=3pc
 \tenrm\baselineskip=12pt
 \noindent
For over 50 years attempts have been made to explain the properties
of nuclear matter in terms of constituent nucleons with very little success.
Here we will investigate one class of many possible models, string-flip
potential models, in which flux-tubes are connected between quarks (in a
gas/plasma) to give a minimal overall field configuration. A general overview
of the current status of these models, along with some of our recently finished
work, shall be given. It shall be shown that these models seem promising in
that they do get most of the bulk properties of nuclear matter correct with the
exception of nuclear binding. Finally we will conclude with a brief discussion
on ways to improve the string-flip potential models in an attempt to obtain
nuclear binding (currently we are investigating short range one-gluon exchange
effects -- some preliminary results shall be mentioned).
\vglue 0.8cm}
%
%
{\bf\noindent 1. Overview}
\vglue 0.4cm
\baselineskip=14pt
The main objective of our work is to attempt to describe nuclear matter in
terms of its constituent quarks. A difficult task indeed, for over the past 50
years many attempts have been made with very little success. The main
difficulty is due to the nonperturbative nature of QCD. The most rigorous
method for handling multiquark systems to date is lattice QCD, but given the
magnitude of our problem it appears unlikely to be useful in the near future,
due to its computationally intensive nature. As a result we must consider more
phenomenological means.

The basic idea here is to construct models which are motivated by lattice QCD
theory and nucleon based models of nuclear matter. A very crude model should be
able to get most of bulk properties of nuclear matter correct: i.e.
\begin{itemize}
\item Nucleon gas at low densities with no van der Waals forces.
\item Nucleon binding at higher densities.
\item Nucleon swelling and saturation of nuclear forces with increasing
density.
\item Quark-gluon plasma at extremely high densities.
\end{itemize}
The are many models out there that attempt to fill this shopping list but
we have found none that covers it completely.

Here, we shall restrict ourselves to a particular class of models, called
string-flip Potential Models\cite{kn:Boyce,kn:HorowitzI,kn:Watson}, which fills
all the items on our list with the exception of nucleon binding. The later part
of our discussion shall be a way of perhaps improving these models in an
attempt to get nucleon binding.
\vglue 0.6cm
{\bf\noindent 2. String-Flip Potential Model}
\vglue 0.4cm
Sting-flip potential models are models which are inspired by lattice QCD in
that they attempt to mimic the flux-tube dynamics.
In the most general setting, one assumes that the quarks, in question, move
slowly enough that their fields have enough time to reconfigure themselves in
order to minimize the overall potential energy: i.e.
\begin{equation}
V=\min
   \{
     \sum_{\{q_m\ldots q_n\}}
     {\rm v}({\vec{\rm r}}_m\ldots {\vec{\rm r}}_n)\,|\,
     \bigcup_{\{m\ldots n\}}^\sim\{q_m\ldots q_n\}=\{q_1\ldots q_{N_q}\}
   \}\;,
\label{eq:cpot}
\end{equation}
where the $N_q$ quarks are placed in a cube of side $L$ and subjected
to periodic boundary conditions, to simulate continuous quark matter.
The sum is over all gauge invariant sets $\{q_m\ldots q_n\}$ of quarks,
such that at least one element from each set lies inside a common box,
whose disjoint union, $\stackrel{\mbox{\footnotesize$\sim$}}{\cup}\,$,
makes up the complete colour singlet set $\{q_1\ldots q_{N_q}\}$ of
$N_q$ quarks. It is easy to see that this potential allows for complete
minimal quark clustering separability at low densities without
suffering from van der Waals forces.

At present these models are quite crude in that they do not include short range
one-gluon exchange phenomena and spin effects, and are flavour degenerate.

\vglue 0.4cm
{\it \noindent 2.1. $SU_{\ell}(3)$ \& $SU_h(3)$ Models}
\vglue 0.1cm
To simplify matters we shall assume that the multiquark potential,
eq.~(\ref{eq:cpot}), runs over triplets of quarks and that the colour is fixed
to a given quark: i.e.
\begin{equation}
V=\min\{\sum_{\{q_rq_gq_b\}}{\rm v}(\vrr,\vrg,\vrb)|\,
        \bigcup_{\{rgb\}}^\sim\{q_rq_gq_b\}=
        \{q_1\ldots q_{N_q}\}
      \}\;.\label{eq:pot}
\end{equation}

For a linear model, $SU_{\ell}(3)$, the three body potential, ${\rm
v}(\vrr,\vrg,\vrb)$, is given by\cite{kn:CarlsonB}
$$
  \setlength{\unitlength}{1in}
  \begin{picture}(6,1.25)
     \put(0,0.625)
        {
         $
          {{\rm v}}_\ell(\vrr,\vrg,\vrb)=\sigma
          \left\{
           \begin{array}{l}
              \rbr+\rrg;\;{\rm if}\;\angle\,brg\,\ge\,120^\circ\\
              \rrg+\rgb;\;{\rm if}\;\angle\,rgb\,\ge\,120^\circ\\
              \rgb+\rbr;\;{\rm if}\;\angle\,gbr\,\ge\,120^\circ\\
              \sqrt{\frac{3}{2}\xi_{\rm rgb}^2+\frac{\sqrt{3}}{2}A};
              \;{\rm otherwise}
           \end{array}
          \right.
         $
        }
     \put(3.25,0.25)
        {
         \mbox{\epsfxsize=5.5cm
	\epsffile{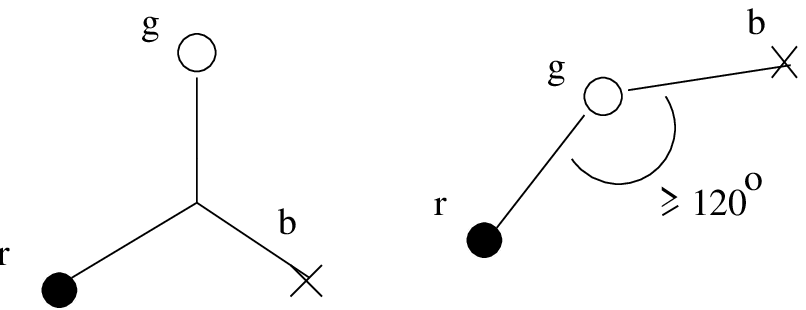}}
        }
     \put(5.78125,0.625){\addtocounter{equation}{1}(\theequation)}
  \end{picture}
$$
where $\vrij=\vri-\vrj\,$, $\xi_{\rm rgb}^2=(\rrg^2+\rgb^2+\rbr^2)/3\,$,
and $A$ is the area enclosed by the triangle $\triangle rgb\,$.

For a harmonic oscillator model, $SU_h(3)$, our three body potential is simply
\begin{equation}
\begin{array}{lr}
{\rm v}_h(\vrr,\vrg,\vrb)=\frac{1}{2}k\xi_{\rm rgb}^2 &
\mbox
 {
  \hspace{0.75cm}
  \raisebox{-0.25cm}
  {
   \mbox{\epsfxsize=1cm\epsffile{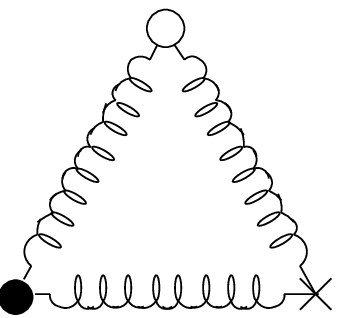}}
  }
 }
\end{array}
\end{equation}
where we have assumed quarks of equal mass ({\small $cf.$ $ref.$}
\raisebox{-1.5mm}{\cite{kn:Boyce}}).

The linear model is inspired by lattice QCD whereas the harmonic oscillator
is used because it has some nice analytical properties which allows us to make
a consistency check.
\vglue 0.4cm
{\it \noindent 2.2. Computation}
\vglue 0.1cm
Here we use a variational procedure to find the binding energy per nucleon,
$E_B$ ($=\bar E(\rho)-\bar E(0)$), and the correlation length,
$\beta$, as a function of nucleon density, $\rho$.

The variational wave function,
\begin{equation}
\Psi_{\alpha\beta\rho}=\chic\Phi_{Slater}\;,\label{eq:vwf}
\end{equation}
is a trial function which consists of the product of a symmetric correlation
piece,
$$
\setlength{\unitlength}{1in}
\begin{picture}(6,1.75)\put(-0.09375,-0.125){
  \setlength{\unitlength}{1in}
  \begin{picture}(6,1.75)
     \put(2.25,1.4375){\vector(1,-2){0.125}}
     \put(1.25,1.625)
        {
         \footnotesize
         \begin{tabular}{c}
            Correlation length\\
            {$\displaystyle\beta\sim\frac{1}{{\rm r}_{\rm rms}}\;.$}
         \end{tabular}
        }
     \put(3.5625,1.625){\vector(-3,-2){0.6875}}
     \put(3.5,1.375)
       {
        \footnotesize
        \begin{tabular}{l}
        Variational parameter\\
        $
         \alpha=\left\{
                       \begin{array}{ll}
                        1.75  & {\rm if}\;{\rm SU}_\ell(3)\\
                        2     & {\rm if}\;{\rm SU}_h(3)
                       \end{array}
                \right.
        $\\
        and is found by\\
        minimizing E($\rho=0$).
        \end{tabular}
       }
     \put(0.75,0.75)
        {
         $
          \chic=
             {\rm e}^{
                      -\frac{1}{2}\;
                      {\displaystyle \sum_{\{{\rm rgb}\}}
                      (\beta\xi_{{\rm rgb}})^\alpha}
                     }
           \;,
         $
        }
     \put(2.3125,0.3625){\vector(-1,2){0.1875}}
     \put(2.25,0.3)
        {
         \footnotesize
         \begin{tabular}{c}
         $\sum$ over the set of triplets\\
         $\{{\rm rgb}\}$ which minimizes $V$.\\
         \end{tabular}
        }
     \put(5.78125,0.75){\addtocounter{equation}{1}(\theequation)}
  \end{picture}}
\end{picture}
$$
and an antisymmetric Fermi piece,
$$
  \setlength{\unitlength}{1in}
  \begin{picture}(6,1.5)
     \put(0.75,1)
        {
         $
          \Phi_{Slater}=
          \phi_{\rm r}(\rho)\phi_{\rm g}(\rho)\phi_{\rm b}(\rho)\;.
         $
        }
     \put(1.75,0.5625)
        {
         \mbox{\epsfxsize=1cm
	\epsffile{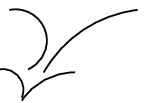}}
        }
     \put(1.0,0.1875)
        {
         \footnotesize
         \begin{tabular}{l}
         \begin{tabular}{c}
            Product of Slater determinants containing\\
            {$\displaystyle\phi_i(\vrj)=\sin(
            \frac{2\pi}{{\rm L}}\,\vn_i\cdot{\vec{{\rm r}}}_j+\delta_i
            )\,,$}\\
         \end{tabular}\\$\;$
         where $\delta_i=0$ or $\pi/2\,$, and
         $(\vn_i)_{a(=x,y,z)}=0,\pm 1,\pm 2,\ldots$
         \end{tabular}
        }
     \put(4.1875,1.6375){\footnotesize Density}
     \put(4,1.375)
        {
         \footnotesize $\displaystyle\rho=\frac{\rm Nucleons}{{\rm L}^3}$
        }
     \put(2.4375,1.1875)
        {
         \mbox{\epsfxsize=4cm
	\epsffile{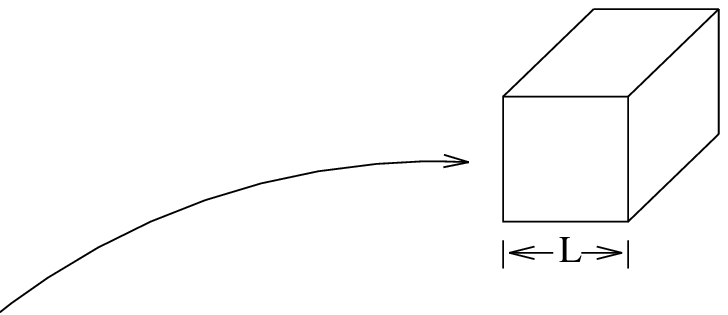}}
        }
     \put(5.78125,1){\addtocounter{equation}{1}(\theequation)}
  \end{picture}
$$
This gives a smooth interpolation between a correlated (nucleon) gas at low
densities and an uncorrelated (Fermi) gas at high densities.

The results of the variational procedure are shown in Fig.~1. The solid
lines show the minimal $E_B(\rho)$ and $\beta(\rho)$ curves, where we have
minimized $\bar E(\rho,\beta)$. The integrals involved in evaluating $\bar
E(\rho,\beta)$ were done using the Metropolis
algorithm\cite{kn:Metropolis,kn:Ceperley}
along with a bag of tricks to minimize cpu time\cite{kn:Boyce}.
\begin{center}
\setlength{\unitlength}{1in}
\begin{picture}(6,3.25)
\put(0,0.4375){\setlength{\unitlength}{1in}
\begin{picture}(6,2.75)
\put(0.5625,2.625){\footnotesize (a)}
\put(3.5625,2.625){\footnotesize (b)}
\put(0,-0.125){\mbox{\epsfxsize=3.25in
	\epsffile{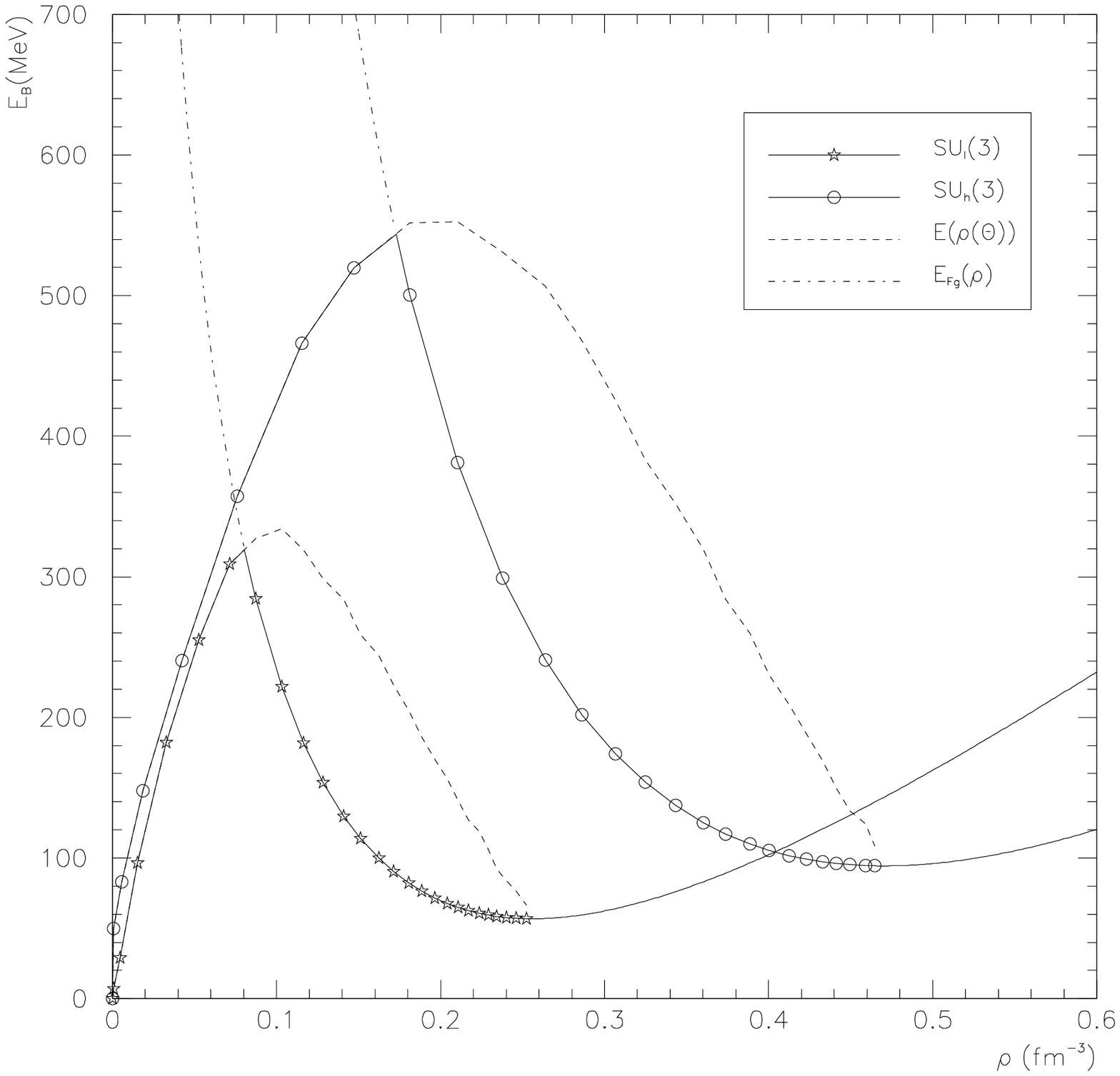}}}
\put(3,-0.125){\mbox{\epsfxsize=3.25in
	\epsffile{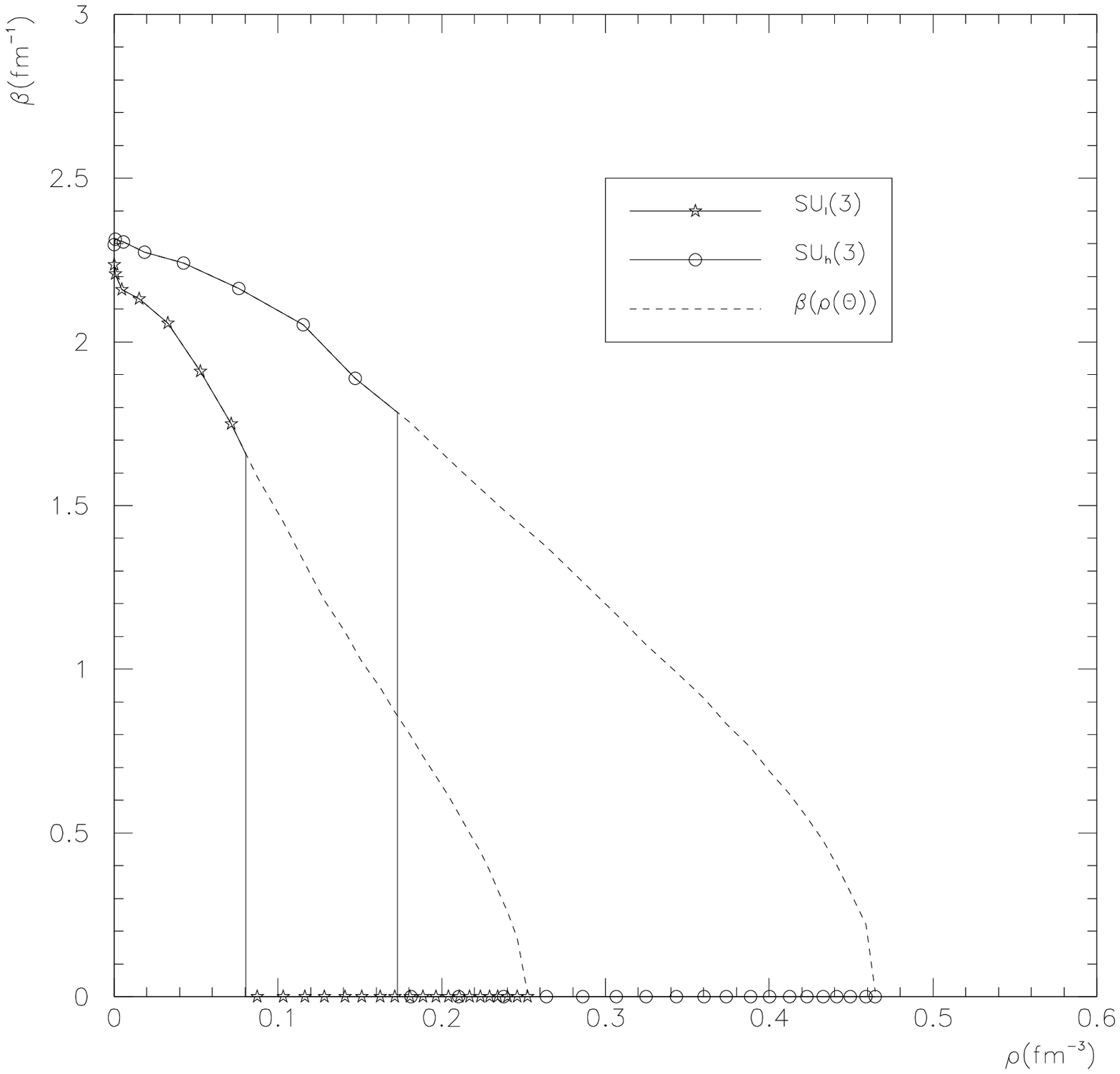}}}
\end{picture}}
\put(0.125,0.125){
\parbox{5.75in}{\footnotesize {\bf Fig. 1:} $E_B(MeV)$, (a), and
$\beta(fm^{-1})$, (b), verses $\rho(fm^{-3})$ for $SU_{\ell}(3)$ ($\star$)
and $SU_h(3)$ ($\circ$), where  $m_q=330MeV/c^2$, $\sigma=0.91GeV/fm$ and
$k\approx3244MeV/fm^2$\cite{kn:Boyce}. The dashed curves represent the
remnants of the minimal $E_B(\rho,0)$ and $E_B(\rho,\beta)$ trajectories.}}
\end{picture}
\end{center}

{}From Fig.~1.a we see that there is an overall saturation of nucleon forces
followed by a transition (the kink in the solid line) to a Fermi gas, but no
nucleon binding. Fig.~1.b illustrates an overall swelling of nucleons with
increasing $\rho$ until they become infinite beyond the transition point.
\vglue 0.6cm
{\bf\noindent 3. The Show So Far!}
\vglue 0.4cm
We have demontrated that string-flip potential models gets the overall bulk
properties of nuclear matter correct with the exception of binding.

\vglue 0.2cm
{\noindent So what can we do to obtain binding?}
\begin{description}
\item [{a)}] What about allowing the colour to move around?\\
      --- This seems unlikely; it has been investigated in an $SU_h(2)$ model
      by HP\cite{kn:HorowitzI}.
\item [{b)}] What about allowing higher order flux-tube topologies?\\
       --- This seems unlikely; it been investigated in an -- albeit slightly
       ill\footnote{Predicts nucleon shrinking\cite{kn:Boyce}.} -- $SU_h(3)$
       chain model by HP\cite{kn:HorowitzI}.
\item [{c)}] What about ``a)'' and ``b)''?\\
      --- This seems unlikely; it has been investigated by many six quark
      (non-string-flip) models\cite{kn:Nzar,kn:Maltman} which suggests
      six quark states like to dissociate into two nucleons.
\item [{d)}] What about including short range forces?\\
      --- This seems likely; we are currently investigating this possibility.
\end{description}
{\it \noindent 3.1. Short Range Forces}
\vglue 0.1cm
At the moment we are investigating the inclusion of colour Coulomb effects. The
difficulty arises here when trying to combine perturbative and nonperturbative
fields in a consistent fashion. To simplify our study we consider an
$SU_\ell(2)$ potential model, which looks a lot like $SU_\ell(3)$
({\small $cf.$
$ref.$} \raisebox{-1.5mm}{\cite{kn:Watson}}), for $q\bar q$ pairs: i.e.
$$
  \setlength{\unitlength}{1in}
  \begin{picture}(6,0.375)
     \put(0.75,0.125)
        {
         $
          {\rm v}_{ij}\sim\left\{
          \begin{array}{ll}
             {\displaystyle\sigma({\rm r}_{ij}-{\rm r}_0)} &
                {\displaystyle{\rm if}\;{\rm r}_{ij}>{\rm r}_0} \\
             {\displaystyle\alpha_s\lambda_{ij}
                \left(\frac{1}{\rm r_{ij}}-\frac{1}{{\rm r}_0}\right)} &
                {\displaystyle\mbox {\rm if}\;{\rm r}_{ij}<{\rm r}_0}
          \end{array}
          \right.
         $
        }
     \put(4,-0.125)
        {
         \mbox{\epsfxsize=2.5cm
	\epsffile{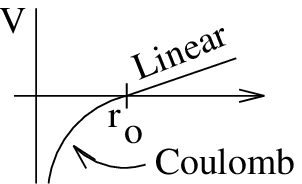}}
        }
     \put(5.78125,0.125){\addtocounter{equation}{1}(\theequation)}
  \end{picture}
$$
where $\lambda_{ij}=-3/4,1/4$ for unlike and like colours respectively, and
$\alpha_s=0.1$. Fig.~2 illustrates the dynamics of the model.
\begin{center}
\framebox{\setlength{\unitlength}{1in}
\begin{picture}(5.75,2.5)\put(-0.125,-0.0625){
\setlength{\unitlength}{1in}
\begin{picture}(5.75,2.5)
\put(0,2){\parbox{3.875in}{\footnotesize {\bf Fig. 2:} Consider configuration
(a) of quarks, with r $>{\rm r}_0$, about to move to (b), s.t. two
of them are within r $<{\rm r}_0$. Then the procedure is to draw a
bubble of ${\rm r}_0$ away from the two, (b), and to cut the flux-tubes at the
boundary and insert virtual $q\bar q$ pairs, (c). Once the potential is
computed the configuration is restored to (b) before the next move is made.}}
\put(0,0.125){\mbox{\epsfysize=2.45in
	\epsffile{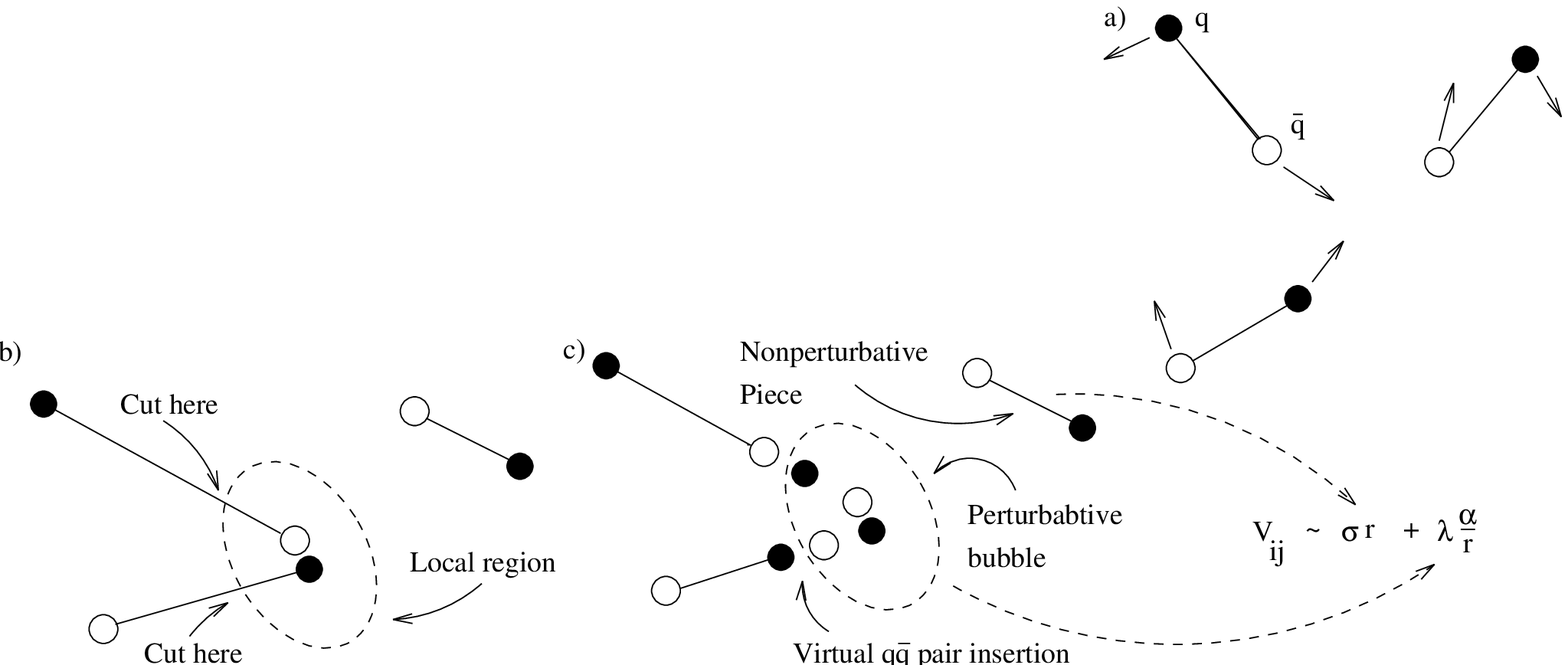}}}
\end{picture}}
\end{picture}}
\end{center}
Notice that this model allows us to construct colourless objects because of the
insertion the virtual $q\bar q$ pairs. These virtual quarks are used as a
tool to get the overall length of flux-tube correct. They are not used in
computing the Coulomb term however, as the field energy is already taken into
account by the ``real'' quarks in the bubble. In general, once the bubbles have
been determined, one must reconfigure the flux-tubes in order to minimize the
linear part of the potential.
\begin{center}
\begin{tabular}{c}
\mbox{}\\
\setlength{\unitlength}{1in}
\begin{picture}(5,1.875)
  \put(0.375,1.75){\small (a)}
  \put(3.3125,1.75){\small (b)}
  \put(0,0){\mbox{\epsfxsize=2.25in
	\epsffile{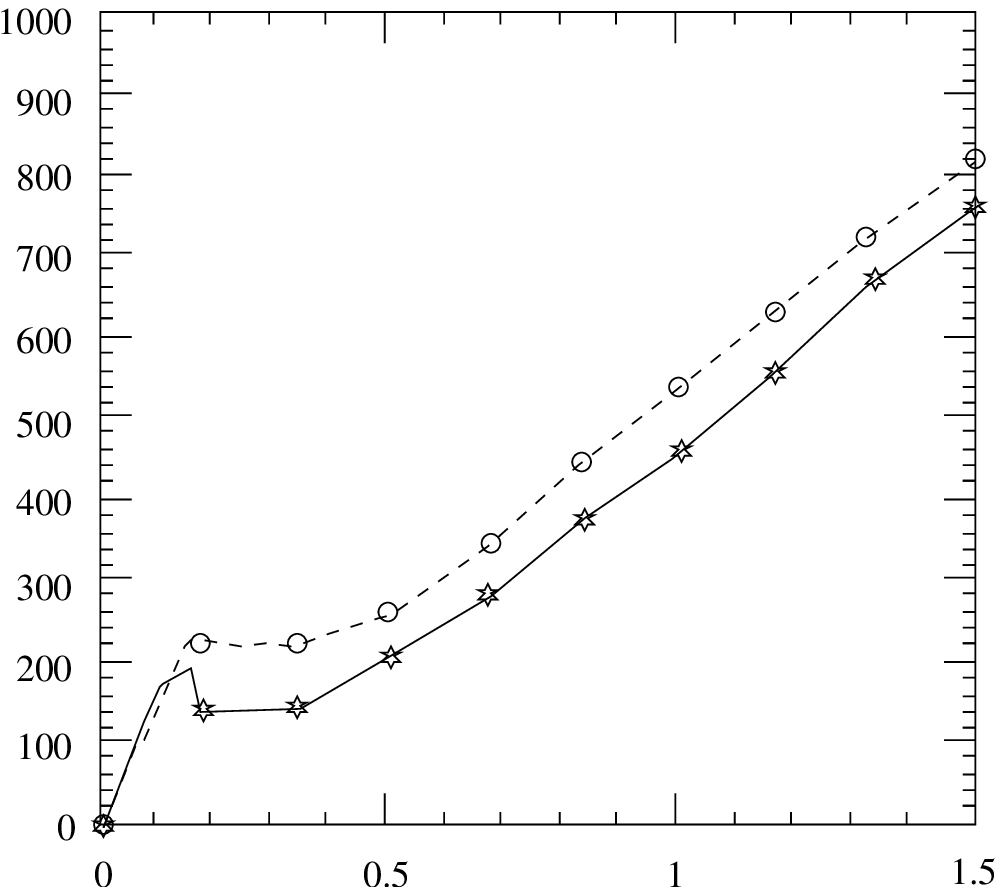}}}
  \put(3,0){\mbox{\epsfxsize=2.16875in
	\epsffile{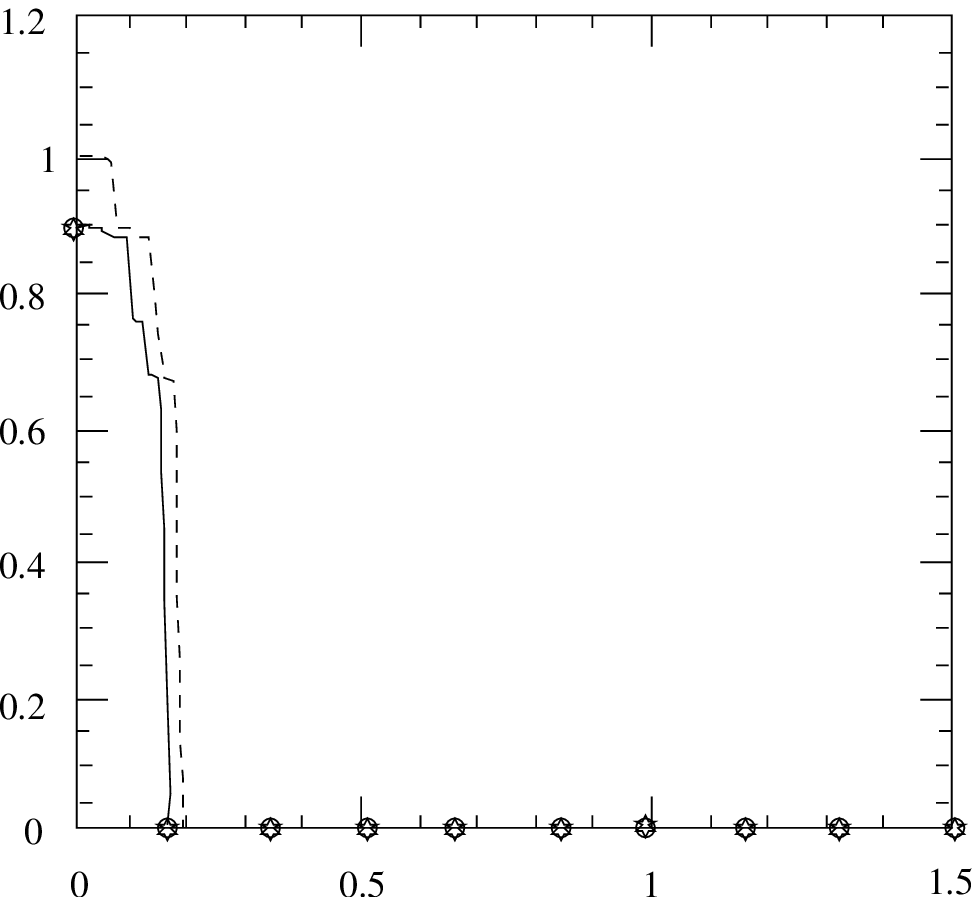}}}
\end{picture}\\
{\footnotesize {\bf Fig. 3:} $E_B(\rho)$, (a), and $\beta(\rho)$, (b), for
${\rm r}_0=0\,$fm ({\tiny$-\,-$}) and ${\rm r}_0=0.1\,$fm (---).}
\end{tabular}
\end{center}
Some preliminary results of these models are shown in Fig.~3. For ${\rm
r}_0=0\,$fm we get back the same results -- as expected -- and for ${\rm
r}_0=0.1\,$fm we get a slightly different answer.

Although the model is currently for $SU_\ell(2)$ it should be easy to extend it
to a full $SU_\ell(3)$ model with all the one-gluon exchange phenomena. The
results presented here are very preliminary as our wave function is slightly
sick. By this we mean that we used our old wave function, eq.~(\ref{eq:vwf}),
where the correlation piece was over the modified flux-tubes and a new piece
was
thrown in to take care of the local Coulomb interactions as they occurred: i.e.
\begin{equation}
\Psi_{\alpha\alpha^\prime\beta\rho}\;\sim
\;\chi^{\rm (Linear)}_{\alpha\beta}\;\times\;
\chi^{\rm (Coulomb)}_{\alpha^\prime\beta}\;\times\;\Phi^{\rm (Fermi)}_\rho\;.
\end{equation}
However because the flux-tubes and bubbles can now be created or destroyed the
wave function is, in general, no longer smooth and continuous: i.e. we are not
guaranteed a variational lower bound. Currently we are working to rectify the
matter and hope to have some results out in the near future.
\vglue 0.6cm
{\bf\noindent 4. Closing Remarks}
\vglue 0.4cm
We have demonstrated that string-flip potential models appear to be fairly
good candidates for explaining nuclear matter at the constituent quark level.
Crude models seem to get most of the bulk properties with the exception of
nuclear binding. We have also shown that it is fairly reasonable to assume that
these models must be extended if we are ever going to achieve binding. In this
light we have modified the model to include one-gluon exchange effects.
Currently it only includes Coulomb effects, which we hope to have results for
sometime soon. Further down the round we would like to include the rest of the
one-gluon exchange phenomena.
\vglue 0.6cm
{\bf\noindent References \hfil}
\vglue 0.4cm

\vglue 0.5cm
\end{document}